\documentclass{article}
 
\usepackage{amsmath,amssymb,amsfonts}   
\usepackage{graphicx,bm}

\usepackage{float}
\usepackage{epsfig}
\usepackage{esint}

\usepackage{bm,braket}

\newcommand{\calU}{{\mathcal U}}

\newcommand{\calP}{{\mathcal P}}
\newcommand{\calQ}{{\mathcal Q}}
\newcommand{\R}{{\mathbb R}}

\newcommand{\X}{\mathbf{X}}
\renewcommand{\P}{\mathbb{P}}
\newcommand{\PP}{\widetilde{P}}
\newcommand{\x}{\mathbf{x}}

\newcommand{\e}{{\mathrm e}}
\newcommand{\E}{{\mathbb E}}

\newcommand{\n}{\mathbf n}
\newcommand{\calT}{{\mathcal T}}
\renewcommand{\P}{\mathbb P}
\newcommand{\p}{\widetilde{p}}

\newcommand{\Q}{\widetilde{Q}}
\newcommand{\ellh}{\hat{\ell}}

\begin{document}

 \title{Diffusion-mediated surface reactions and stochastic resetting}

\author{\em 
Paul C. Bressloff \\ Department of Mathematics, University of Utah \\
155 South 1400 East, Salt Lake City, UT 84112}
\maketitle

\begin{abstract} 
In this paper, we investigate the effects of stochastic resetting on diffusion in $\R^d\backslash \calU$, where $\calU$ is a bounded obstacle with a partially absorbing surface $\partial \calU$. We begin by considering a Robin boundary condition with a constant reactivity $\kappa_0$, and show how previous results are recovered in the limits $\kappa_0\rightarrow 0,\infty$. We then generalize the Robin boundary condition to a more general probabilistic model of diffusion-mediated surface reactions using an encounter-based approach. The latter considers the joint probability density or propagator $P(\x,\ell,t|\x_0)$ for the pair $(\X_t,\ell_t)$ in the case of a perfectly reflecting surface, where $\X_t$ and $\ell_t$ denote the particle position and local time, respectively. The local time determines the amount of time that a Brownian particle spends in a neighborhood of the boundary. The effects of surface reactions are then incorporated via an appropriate stopping condition for the boundary local time. We construct the boundary value problem (BVP) satisfied by the propagator in the presence of resetting, and use this to derive implicit equations for the marginal density of particle position and the survival probability. We highlight the fact that these equations are difficult to solve in the case of non-constant reactivities, since resetting is not governed by a renewal process. We then consider a simpler problem in which both the position and local time are reset. In this case, the survival probability with resetting can be expressed in terms of the survival probability without resetting, which allows us to explore the dependence of the MFPT on the resetting rate $r$ and the type of surface reactions. The theory is illustrated using the example of a spherically symmetric surface.

\end{abstract}

\section{Introduction}

\begin{figure}[b!]
\raggedleft
\includegraphics[width=6cm]{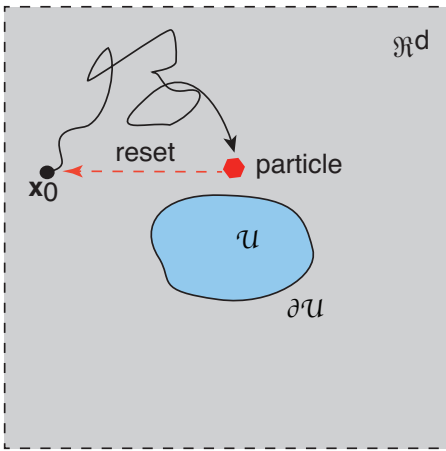}
\caption{Schematic illustration of a particle diffusing in the domain $\R^d\backslash \calU$, where $\calU$ is an obstacle or target with a fixed boundary condition on the surface $\partial \calU$. The particle resets to its initial position $\x_0$ at a constant rate $r>0$. If the surface is totally reflecting, then the probability density converges to an NESS, whereas if the surface is totally absorbing then the particle is eventually absorbed with probability one and the MFPT $T(\x_0)$ is a unimodal function of the resetting rate.}
\label{fig1}
\end{figure}

In recent years there has been a rapid growth of interest in stochastic processes with resetting. A canonical example is a Brownian particle whose position is reset randomly in time at a constant rate $r$ (Poissonian resetting) to some fixed point $x_0$, which is usually identified with its initial position \cite{Evans11a,Evans11b,Evans14}. Consider, in particular, the scenario shown in Fig. \ref{fig1} where there is some obstacle or target $\calU$ located around the origin in $\R^d$. The behavior of the particle will depend on the rate of resetting $r$, the boundary condition on the surface $\partial \calU$ (which is assumed to be smooth), and the dimension $d$. First suppose that there is no resetting ($r=0$). In the case of a totally reflecting boundary $\calU$, the probability density $p(\x,t|\x_0)\rightarrow 0$ as $t\rightarrow \infty$ pointwise, but the survival probability $Q(\x_0,t)\equiv\int_{\R^d\backslash \calU}p(\x,t|\x_0)d\x=1$ for all $t$. On the other hand, if the boundary $\calU$ is totally absorbing, then in the limit $t\rightarrow \infty$ we have $p(\x,t|\x_0) \rightarrow 0$ for all $d\geq 1$, whereas $Q(\x_0,t) \rightarrow 0$ for $d\leq 2$ (recurrence) and $Q(\x_0,t) \rightarrow Q_{\infty}>0$ for $d>2$ (transience). Irrespective of the dimensionality, the corresponding mean first passage time(MFPT) $T(\x_0)=\int_0^{\infty}Q(\x_0,t)dt$ is infinite. Now suppose that resetting occurs at a fixed rate $r>0$. If the surface is totally reflecting, then the probability density converges to a nonequilibrium stationary state (NESS) \cite{Evans11a,Evans11b,Evans14,Majumdar15}. On the other hand, if the surface is totally absorbing then the MFPT $T(\x_0)$ is finite and typically has an optimal value as a function of the resetting rate $r$ \cite{Evans11a,Evans11b,Evans14}. These results carry over to more general stochastic processes with resetting, including non-diffusive processes such as Levy flights \cite{Kus14} and active run and tumble particles \cite{Evans18,Bressloff20}, diffusion in switching environments \cite{Bressloff20a,Bressloff20b} or potential landscapes \cite{Pal15}, resetting followed by a refractory period \cite{Evans19a,Mendez19a}, and resetting with finite return times \cite{Pal19a,Mendez19,Bodrova20,Pal20,Bressloff20c}. (For further generalizations and applications see the review \cite{Evans20} and references therein.)

In this paper, we extend the problem shown in Fig. \ref{fig1} to the case of a partially absorbing surface $\partial \calU$. We begin by considering the simplest case of a Robin boundary condition with a constant reactivity $\kappa_0$ (section 2). We proceed by solving the forward diffusion equation in Laplace space, which allows us to relate the survival probability and MFPT to the corresponding quantities without resetting. We explore the behavior of the MFPT as a function of the parameters $\kappa_0$ and $r$, and show how previous results are recovered in the limits $\kappa_0\rightarrow 0$ and $\kappa_0\rightarrow \infty$. For the sake of illustration, we consider the examples of diffusion on the half-line and its higher-dimensional analog, namely, a spherically symmetric obstacle. 

In section 3, we generalize the Robin boundary condition to a more general probabilistic model of diffusion-mediated surface reactions, following the encounter-based approach developed by Grebenkov \cite{Grebenkov19b,Grebenkov20,Grebenkov21}. The latter exploits the fact that diffusion in a domain with a totally reflecting surface can be implemented probabilistically in terms of so-called reflected Brownian motion, which involves the introduction of a Brownian functional known as the boundary local time \cite{Levy39,McKean75,Majumdar05}. The local time characterizes the amount of time that a Brownian particle spends in the neighborhood of a point on the boundary. In the encounter-based approach, one considers the joint probability density or propagator $P(\x,\ell,t|\x_0)$ for the pair $(\X_t,\ell_t)$ in the case of a perfectly reflecting surface, where $\X_t$ and $\ell_t$ denote the particle position and local time, respectively. The effects of surface reactions are then incorporated via an appropriate stopping condition for the boundary local time. The propagator satisfies a corresponding boundary value problem (BVP), which can be derived using integral representations \cite{Grebenkov20} or path integrals \cite{Bressloff22}. Here we show how to incorporate stochastic resetting into the propagator BVP and use this to derive corresponding implicit equations for the marginal density of particle position and the survival probability. However, these equations are difficult to solve due to the fact that resetting is no longer governed by a renewal process. Therefore, we consider a simpler problem in which both the position and local time are reset, in which case the survival probability with resetting can be expressed in terms of the survival probability without resetting. This allows us to explore the dependence of the MFPT on the resetting rate $r$ and the type of surface reactions. The theory is illustrated using the example of a spherically symmetric surface. In addition to determining the optimal resetting rate that minimizes the MFPT, we also show that the relative increase in the MFPT compared to the case of a totally absorbing surface can itself exhibit non-monotonic variation with $r$.

\setcounter{equation}{0}
\section{Diffusion with stochastic resetting and a partially reflecting boundary}

\subsection{Diffusion on the half-line}

Consider a particle diffusing along the half-line $x\in [0,\infty)$ with a partially reflecting boundary at $x=0$.
The probability density $p(x,t|x_0)$ evolves according to the equation \cite{Evans11a,Evans11b}
\begin{subequations}
\begin{align}
\label{1Da}
 &\frac{\partial p}{\partial t}=D\frac{\partial^2p}{\partial x^2} -rp+r Q_r(x_0,t)\delta(x-x_0),\quad x>0,\\
&D\frac{\partial p}{\partial x}+\kappa_0p=0,\quad x=0,\quad
 p(x,0|x_0) = \delta(x - x_0).
\label{1Db}
\end{align}
\end{subequations}
We have introduced the marginal distribution
\begin{equation}
\label{1DQ}
Q_r(x_0,t)=\int_0^{\infty} p(x,t|x_0)dx,
\end{equation}
which is the survival probability that the particle hasn't been absorbed at $x=0$ in the time interval $[0,t]$, having started at $x_0$. The $r$ subscript indicates that $Q_r$ is the survival probability in the presence of resetting
at a rate $r$.
Laplace transforming equations (\ref{1Da}) and (\ref{1Db}) gives
\begin{subequations}
\begin{align}
\label{1DLTa}
  &D\frac{\partial^2\p(x,s|x_0)}{\partial x^2} -(r+s)p(x,s|x_0)=-[1+r \Q_r(x_0,s)]\delta(x-x_0),\quad x>0,\\
 &D\frac{\partial \p(x,s|x_0)}{\partial x}=\kappa_0\p(x,s|x_0),\quad x=0.
\label{1DLTb}
\end{align}
\end{subequations}

The general solution of equation (\ref{1DLTa}) is of the form
\begin{equation}
\label{1Dgir}
\p(x,s|x_0)=A_r(s)\e^{-\alpha x}+[1+r\Q_r(x_0,s)]G(x,\alpha|x_0),
\end{equation}
where $\alpha=\sqrt{[r+s]/D}$.
The first term on the right-hand side of equation (\ref{1Dgir}) is the solution to the homogeneous version of equation (\ref{1Da}) and $G$ is the modified Helmholtz Green's function in the case of a totally absorbing boundary condition at $x=0$:
  \begin{subequations}
\begin{align}
\label{Ga1D}
&D\frac{\partial^2G}{\partial \rho^2}  -(r+s)G  = -\delta(x - x_0), \ 0<x<\infty,\\ 
 & G(0,\alpha|x_0)=0.
 \label{Gb1D}
\end{align}
\end{subequations}
The latter is given by
\begin{align}
\label{GG1D}
 G(x, \alpha|x_0) = \frac{1}{2D\alpha }\left [\e^{-\alpha |x-x_0|}-\e^{-\alpha |x+x_0|}\right ].
 \end{align}
 
 The unknown coefficient $A_r(s)$ is determined by imposing the Robin boundary condition (\ref{1DLTb}):
 \begin{align}
 A_r(s)=\frac{1+r\Q_r(x_0,s)}{\kappa_0 +\alpha D}\e^{-\alpha x_0}.
 \end{align}
 Hence, the full solution of the Laplace transformed probability density with resetting is
  \begin{align}
\label{p1D}
   \p(x, s|x_0) =  [1+r\Q_r(x_0,s)]\p_0(x,r+s|x_0), \quad 0<x<\infty,
\end{align}
where $\p_0$ is the corresponding solution without resetting,
 \begin{equation}
 \p_0(x,s|x_0)=\frac{\e^{-\sqrt{s/D} (x+x_0)}}{\kappa_0+\sqrt{sD}}+G(x, \sqrt{s/D}|x_0).
 \end{equation}
 Finally, Laplace transforming equation (\ref{1DQ}) and using (\ref{p1D}) shows that
 \begin{align}
  \Q_r(x_0,s)&=\int_0^{\infty}\p(x,s|x_0)dx= [1+r\Q_r(x_0,s)]\int_0^{\infty} \p_0(x,r+s|x_0)dx\nonumber \\
&=[1+r\Q_r(x_0,s)]\Q_0(x_0,r+s),
\label{QQr}
 \end{align}
 where $\Q_0$ is the Laplace transform of the survival probability without resetting:
 \begin{equation}
 \Q_0(x_0,s)=\frac{1-\e^{-\sqrt{s/D}x_0}}{s}+\frac{\e^{-\sqrt{s/D}x_0}}{s+\kappa_0\sqrt{s/D}}.
 \end{equation}
  Rearranging equation (\ref{QQr}) thus determines the survival probability with resetting in terms of the corresponding probability without resetting:
 \begin{equation}
 \label{Qr}
 \Q_r(x_0,s)=\frac{\Q_0(x_0,r+s)}{1-r\Q_0(x_0,r+s)}.
 \end{equation}
 This particular relation is identical to one previously derived for totally absorbing surfaces \cite{Evans11a,Evans20}. In the latter case, equation (\ref{Qr}) follows naturally from renewal theory. As we show in this paper, an equation of the form (\ref{Qr}) holds for a general class of diffusion-mediated surface reactions, even though resetting is no longer given by a renewal process. Intuitively speaking, renewal theory does not apply because the amount of time the particle interacts with the partially reactive surface is not reset. (Within the encounter-based framework this can be understood in terms of the boundary local time, see section 3.) On the other hand, in the case of a totally absorbing target, the process is stopped on first encounter with the surface and thus resetting is governed by a renewal process.
 
 \begin{figure}[b!]
\raggedleft
\includegraphics[width=12cm]{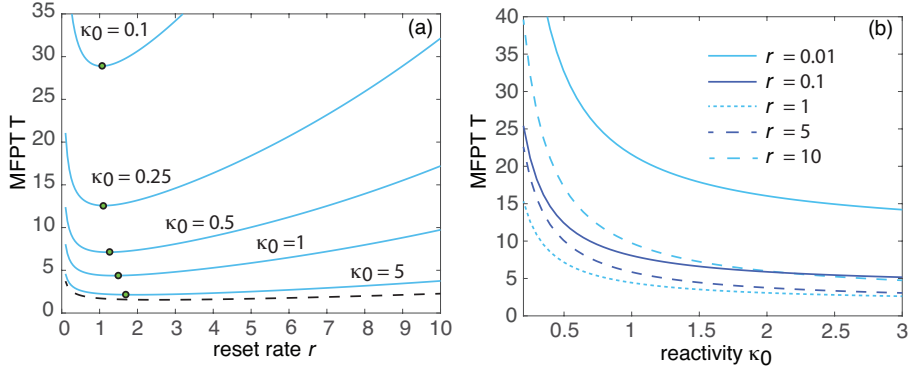}
\caption{Particle diffusing on the half line $x\in [0,\infty)$ with a Robin boundary condition at $x=0$. The particle resets to its initial position $x_0$ at a constant rate $r>0$. (a) Plot of the MFPT $T(x_0)$ as a function of the resetting rate $r$ for various reactivities $\kappa_0$. The dashed line indicates the case of a totally absorbing boundary ($\kappa_0\rightarrow \infty$). (b) Corresponding plots as a function of $\kappa_0$ for various $r$. Other parameter values are $D=1$ and $x_0=1$. Green dots indicate the optimal resetting rates. }
\label{fig2}
\end{figure}

 For $\kappa_0>0$ we expect the steady-state survival probability to vanish, both without and with resetting, since 1D diffusion is recurrent so that absorption eventually occurs. Indeed,
 \begin{align}
 Q_r^*(x_0)&=\lim_{s\rightarrow 0}s\Q_r(x,s|x_0) =\lim_{s\rightarrow 0}\frac{s\Q_0(x_0,r)}{1-r\Q_0(x_0,r)}=0.
 \end{align}
 We have used the fact that $\Q_0(x_0,r)\neq 1/r$ when $\kappa_0>0$. Since $f_r(x_0,t)=-\partial_tQ_r(x_0,t)$ is the first passage time (FPT) density for absorption at $x=0$, it follows that the mean FPT (MFPT) is
 \begin{equation}
  T_r(x_0)=\int_0^{\infty}tf_r(x_0,t)dt=-\int_0^{\infty}t\partial_tQ_r(x_0,t)dt=\int_0^{\infty}Q_r(x_0,t)dt=\Q_r(x_0,0),
 \end{equation}
 where we have used integration by parts. Hence, setting $s=0$ in equation (\ref{Qr}) recovers another well-known relation, namely,
 \begin{equation}
 T_r(x_0)=\frac{\Q_0(x_0,r)}{1-r\Q_0(x_0,r)}.
 \end{equation}
 On the other hand, if $\kappa_0=0$ (totally reflecting boundary at $x=0$), then $\Q_r(x_0,s) =1/s $ for all $x_0<\infty$ and thus $Q_r^*(x_0)=1$. In this special case, there exists a non-trivial stationary state (NESS) given by
 \begin{align}
 p^*(x)&=\lim_{s\rightarrow 0}s\p(x,s|x_0)= \lim_{s\rightarrow 0}s[1+r\Q_r(x_0,s)]\p_0(x,r+s|x_0)\nonumber \\
 &=r\p_0(x,r|x_0)=\frac{1}{2}\sqrt{\frac{r}{D}}\left [\e^{-\sqrt{r/D} |x-x_0|}+\e^{-\sqrt{r/D} |x+x_0|}\right ],
 \end{align}
 which recovers the well-known result of Refs. \cite{Evans11a,Evans11b}. Finally, consider the limit $\kappa_0\rightarrow \infty$ (totally absorbing boundary at $x=0$). In this case,
 \begin{equation}
 \Q_0(x_0,s)=\frac{1-\e^{-\sqrt{s/D}x_0}}{s}
  \end{equation}
  so that
  \begin{equation}
  T_r(x_0)=\frac{1}{r}\left [\e^{\sqrt{r/D}x_0}-1\right ].
  \end{equation}
  In Fig. \ref{fig2}(a) we show sample plots of the MFPT $T_r$ as a function of the resetting rate $r$ for various reactivities $\kappa_0$. As expected, $T_r$ is a unimodal function of $r$ with an optimal resetting rate $r_{\rm opt}$ that is a weakly increasing function of $\kappa_0$. Corresponding plots of $T_r$ as a function of $\kappa_0$ are shown in Fig. \ref{fig2}(b). For fixed $r$, the MFPT is a monotonically decreasing of $\kappa_0$ with a horizontal asymptote given by the MFPT for a totally absorbing target. On the other hand, $T_r$ blows up as $\kappa_0\rightarrow 0$, which indicates the approach to a totally reflecting surface.

\subsection{Diffusion in $\R^d$ with a partially reflecting spherical obstacle at the origin.}

Consider a particle diffusing in ${\mathcal M}\equiv \R^d\backslash \calU$, where $\calU$ is a sphere of radius $\rho_1$ centered at the origin. Suppose that the spherical surface $\partial \calU$ is partially absorbing with constant reactivity $\kappa_0$. Following \cite{Redner01}, the initial position of the particle is randomly chosen from the surface of the sphere $\calU_0$ of radius $\rho_0$, $\rho_1<\rho_0<\infty$. This will allow us to exploit spherical symmetry such that $p=p(\rho,t|\rho_0)$. Finally, we assume that the particle can reset to a random point on the initial spherical surface $\partial \calU_0$ at a Poisson rate $r$.
Setting $\rho=|\x|$, It follows that the density $p$ evolves according to the equation
\begin{subequations}
\begin{align}
\label{mastera}
  &\frac{\partial p}{\partial t}=D\frac{\partial^2p}{\partial \rho^2} + D\frac{d - 1}{\rho}\frac{\partial p}{\partial \rho} -rp+r\Gamma_dQ_r(\rho_0,t)\delta(\rho-\rho_0),\quad \rho_1<\rho ,\\
  &D\frac{\partial p}{\partial \rho}=\kappa_0p,\quad \rho=\rho_1,\quad
 p(\rho,0|\rho_0) =\Gamma_d\delta(\rho - \rho_0),\ \Gamma_d=\frac{1}{\Omega_d \rho_0^{d - 1}},
\label{masterb}
\end{align}
\end{subequations}
where $\Omega_d$ is the surface area of a unit sphere in $\mathbb{R}^d$ and $Q_r(\x_0,t)$ is the survival probability that the particle hasn't been absorbed by the surface $\partial \calU$ in the time interval $[0,t]$, having started at $\rho_0$:
\begin{equation}
\label{S1}
Q_r(\rho_0,t)=\int_{\R^d\backslash \calU} p(\x,t|\x_0) d\x =\Omega_d \int_{\rho_1}^{\infty}\rho^{d-1}p(\rho,t|\rho_0)d\rho.
\end{equation} 
Differentiating both sides of equation (\ref{S1}) with respect to $t$ and using the diffusion equation implies that
\begin{align}
 \frac{\partial Q_r(\rho_0,t)}{\partial t}&=D\Omega_d \int_{\rho_1}^{\infty} \frac{d}{d\rho}\rho^{d-1}\frac{d p(\rho,t|\rho_0)}{d\rho} d\rho-r Q_r(\rho_0,t)+rQ_r(\rho_0,t) \nonumber \\
 &=-J_r(\rho_0,t),\label{Q2}
\end{align}
where $J_r(\rho_0,t)$ is the total probability flux into the surface $\partial \calU$ at time $t$:
\begin{equation}
\label{Jsph}
{J}_r(\rho_0,t)=D\Omega_d\rho_1^{d-1}\left . \frac{\partial  p(\rho,t|\rho_0)}{\partial \rho}\right |_{\rho=\rho_1}.
\end{equation}

Laplace transforming equation (\ref{mastera}) and (\ref{masterb}) gives
\begin{subequations}
\begin{align}
\label{spha}
 &D\frac{\partial^2\p(\rho,s|\rho_0)}{\partial \rho^2} + D\frac{d - 1}{\rho}\frac{\partial \p(\rho,s|\rho_0)}{\partial \rho}-(r+s)\p(\rho,s|\rho_0)\\
 &\hspace{5cm} =-[1+r \Q_r(\rho_0,s)]\Gamma_d\delta(\rho-\rho_0) ,\quad \rho_1<\rho,\nonumber \\
  &D\frac{\partial \p(\rho,s|\rho_0)}{\partial \rho}=\kappa_0 \p(\rho,s|\rho_0) ,\quad \rho=\rho_1.
\label{sphb}
\end{align}
\end{subequations}
Equations of the form (\ref{spha}) can be solved in terms of modified Bessel functions \cite{Redner01}. The general solution is
  \begin{align}
\label{qir}
    \p(\rho, s|\rho_0) = A_r(s)\rho^\nu K_\nu(\alpha \rho) + [1+r\Q_r(\rho_0,s)]G(\rho, \alpha| \rho_0), \ \rho_1<\rho,
\end{align}
where $\nu = 1 - d/2$, $\alpha=\sqrt{[r+s]/D}$,  and $K_{\nu}$ is a modified Bessel function of the second kind.
The first term on the right-hand side of equation (\ref{qir}) is the solution to the homogeneous version of equation (\ref{spha}) and $G$ is the modified Helmholtz Green's function in the case of a totally absorbing surface $\partial \calU$:
  \begin{subequations}
\begin{align}
\label{Ga}
&D\frac{\partial^2G}{\partial \rho^2} + D\frac{d - 1}{\rho}\frac{\partial G}{\partial \rho} -(r+s)G  = -\Gamma_d\delta(\rho - \rho_0), \ \rho_1<\rho,\\ 
 & G(\rho_1,\alpha|\rho_0)=0.
 \label{Gb}
\end{align}
\end{subequations}
The latter is given by \cite{Redner01}
\begin{align}
\label{GGs}
   G(\rho, s| \rho_0) = \frac{ (\rho\rho_0)^\nu }{D\Omega_d}\frac{[I_{\nu}(\alpha \rho_<)K_{\nu}(\alpha \rho_1)-I_{\nu}(\alpha \rho_1)K_{\nu}(\alpha \rho_<)]K_{\nu}(\alpha \rho_>)}{K_{\nu}(\alpha \rho_1)},
\end{align}
where $\rho_< = \min{(\rho, \rho_0)}$, $\rho_> = \max{(\rho, \rho_0)}$, and $I_{\nu}$ is a modified Bessel function of the first kind.
The unknown coefficient $A_r(s)$ is determined from the boundary condition (\ref{sphb}):
\begin{align}
  \kappa_0 A_r(s)F_{\alpha}(\rho_1) &=DA_r (s) F_{\alpha}'(\rho_1)+D[1+r\Q_r(\rho_0,s)]\left . \frac{d}{d\rho}G(\rho,\alpha|\rho_0) \right |_{\rho=\rho_1},
\label{CB3}
\end{align}
with
\begin{align}
\label{boll}
\left . D\frac{d}{d\rho}G(\rho,\alpha|\rho_0) \right |_{\rho=\rho_1}=\frac{1}{\Omega_d\rho_1^{d-1}} \frac{F_{\alpha}(\rho_0)}{F_{\alpha}(\rho_1)}..
\end{align}
We have set
\begin{equation}
\label{FF}
F_{\alpha}(\rho)=\rho^{\nu}K_{\nu}(\alpha \rho),\quad F_{\alpha}'(\rho)=\nu \rho_1^{\nu-1} K_\nu(\alpha \rho_1) +\alpha \rho_1^{\nu}K'_{\nu}(\alpha \rho_1).
\end{equation}
Rearranging (\ref{CB3}) shows that 
\begin{align}
\label{AA}
A_r(s)=\frac{D [1+r\Q_r(\rho_0,s)]}{\kappa_0F_{\alpha}(\rho_1)-DF_{\alpha}'(\rho_1)} \left . \frac{d}{d\rho}G(\rho,\alpha|\rho_0) \right |_{\rho=\rho_1}.
\end{align}
Hence, the full solution of the Laplace transformed probability density with resetting is
  \begin{align}
\label{pr}
   \p(\rho, s|\rho_0) =  [1+r\Q_r(\rho_0,s)]\p_0(\rho,r+s|\rho_0), \ \rho_1<\rho,
\end{align}
where $\p_0$ is the corresponding solution without resetting,
\begin{align}
\label{pr0}
    \p_0(\rho, s|\rho_0) =A_0(s) \rho^\nu K_\nu(\sqrt{s/D} \rho) + G(\rho, \sqrt{s/D}| \rho_0), \ \rho_1<\rho.
\end{align}

Multiplying both sides of equation (\ref{pr}) by $\Omega_d\rho^{d-1}$, integrating with respect to $\rho$ and using equation (\ref{S1}) implies that
\begin{align}
\label{Qr0}
   \Q_r(\rho_0, s) =  [1+r\Q_r(\rho_0,s)]\Q_0(\rho_0,r+s),
\end{align}
where $\Q_0$ is the corresponding survival probability without resetting. Rearranging this equation yields the higher-dimensional analog of equation (\ref{Qr})
\begin{equation}
\label{Qrel}
\Q_r(\rho_0,s)=\frac{\Q_0(\rho_0,r+s)}{1-r\Q_0(\rho_0,r+s)}.
\end{equation}
Moreover, Laplace transforming equation (\ref{Q2}) and noting that $Q_r(\rho_0,0)=1$ gives
\begin{equation}
\label{QL}
 s\widetilde{Q}_r(\rho_0,s)-1=- \widetilde{J}_r(\rho_0,s)=- [1+r\Q_r(\rho_0,s)]\widetilde{J}_0(\rho_0,r+s),
\end{equation}
where $\widetilde{J}_0$ is the flux without resetting. We have used equation (\ref{pr}). Rearranging equation (\ref{QL}) then gives
\begin{equation}
\label{QF}
\Q_r(\rho_0,s)=\frac{1-\widetilde{J}_0(\rho_0,r+s)}{s+r\widetilde{J}_0(\rho_0,r+s)}.
\end{equation}

Let ${\mathcal T}_r$ denote the first passage time
\begin{equation}
\label{Tell}
{\mathcal T}_r=\inf\{t>0,\,|\X(t)|=\rho_1\},
\end{equation}
where $\X(t)$ is the position of the particle at time $t$. Since
\begin{equation}
Q_r(\rho_0,t)=\P[{\mathcal T}_r>t,\, |\X(0)|=\rho_0],
\end{equation}
it follows that the probability density of the first passage time $\calT$ is $-\partial Q/\partial t$, and the MFPT is 
\begin{align}
\label{MFPT1}
T_r(\rho_0)&=-\int_0^{\infty}t\frac{\partial Q_r(\rho_0,t)}{\partial t}dt =\int_0^{\infty} Q_r(\rho_0,t)dt = \widetilde{Q}_r(\rho_0,0).
\end{align}
We have used integration by parts. Finally, taking the limit $s\rightarrow 0$ in equation (\ref{Qrel}) gives
\begin{equation}
T_r(\rho_0)=\frac{\Q_0(\rho_0,r)}{1-r\Q_0(\rho_0,r)}.
\end{equation}
We will explore the behavior of $T_r$ as a function of $r$ and $\gamma =\kappa_0/D$ in section 4,  where we consider more general surface reaction schemes. Here we consider the case of a totally absorbing surface. In the limit $\kappa_0\rightarrow \infty$, we have
\begin{subequations}
\begin{align}
\label{pinf}
  \p_{0}(\rho,s|\rho_0)&\longrightarrow G(\rho,\sqrt{s/D}|\rho_0),  \\ 
 \widetilde{J}_{0}(\rho_0,r+s)&\equiv D\Omega_d\rho_1^{d-1}\left . \frac{\partial  \p_{0}(\rho,r+s|\rho_0)}{\partial \rho}\right |_{\rho=\rho_1}\longrightarrow \frac{\rho_0^{\nu}K_{\nu}(\alpha \rho_0)}{\rho_1^{\nu}K_{\nu}(\alpha \rho_1)}.
\label{Jinf}
\end{align}
\end{subequations}
Substituting into equation (\ref{QF}) gives
\begin{equation}
\label{QFinf}
\Q_r(\rho,s)=\frac{ \rho_1^{\nu}K_{\nu}(\alpha \rho_1)-\rho_0^{\nu}K_{\nu}(\alpha \rho_0)}{s\rho_1^{\nu}K_{\nu}(\alpha \rho_1)+r\rho_0^{\nu}K_{\nu}(\alpha \rho_0)}
\end{equation}
and, hence,
\begin{equation}
\label{Tinf}
T_r(\rho_0)=\frac{1}{r}\left [ \frac{\rho_1^{\nu}K_{\nu}(\alpha \rho_1)}{\rho_0^{\nu}K_{\nu}(\alpha \rho_0)}-1\right ].
\end{equation}
Equations (\ref{QFinf}) and (\ref{Tinf}) recover the results for a totally absorbing spherical surface obtained in Ref. \cite{Evans14}.

\setcounter{equation}{0}
  \section{Diffusion-mediated surface reactions}
  
  \subsection{Diffusion-mediated surface reactions without resetting}

Recently, it has been shown how to reformulate the Robin boundary condition for a diffusing particle without resetting using a probabilistic interpretation based on the so-called boundary local time \cite{Grebenkov19b,Grebenkov20,Grebenkov21,Bressloff22}. The latter is a Brownian functional that keeps track of the amount of time a particle spends in a local neighborhood of a boundary. Let $\X_t \in\R^d\backslash \calU$ represent the position of a diffusing particle at time $t$ with an obstacle ${\mathcal U}$ centered at the origin. If the surface $\partial \calU$ is totally reflecting then we can define a boundary local time according to \cite{Grebenkov19a}
\begin{equation}
\label{loc}
\ell_t=\lim_{h\rightarrow 0} \frac{D}{h} \int_0^t\Theta(h-\mbox{dist}(\X_{\tau},\partial \calU))d\tau,
\end{equation}
where $\Theta$ is the Heaviside function. Note that $\ell_t$ has units of length due to the additional factor of $D$. Let $P_0(\x,\ell,t|\x_0)$ denote the joint probability density or propagator for the pair $(\X_t,\ell_t)$ and introduce the stopping time \cite{Grebenkov19b,Grebenkov20,Grebenkov21}
\begin{equation}
\label{Tell0}
{\mathcal T}_0=\inf\{t>0:\ \ell_t >\widehat{\ell}\},
\end{equation}
 with $\widehat{\ell}$ an exponentially distributed random variable that represents a stopping local time. That is, $\P[\widehat{\ell}>\ell]=\e^{-\gamma\ell}$
with $\gamma=\xi^{-1}=\kappa_0/D$. (Roughly speaking, the stopping time ${\mathcal T}_0$ is a random variable that specifies the time of absorption, which is determined by the instant at which the local time $\ell_t$ crosses a random threshold  $\ellh$.) The relationship between $p_0(\x,t|\x_0)$ and $P_0(\x,\ell,t|\x_0)$ can then be established by noting that
\[p_0(\x,t|\x_0)d\x=\P_0[\X_t \in (\x,\x+d\x), \ t < {\mathcal T}_0|\X_0=\x_0].\]
Given that $\ell_t$ is a nondecreasing process, the condition $t < {\mathcal T}_0$ is equivalent to the condition $\ell_t <\widehat{\ell}$. This implies that \cite{Grebenkov20}
\begin{align*}
p_0(\x,t|\x_0)d\x&=\P[\X_t \in (\x,\x+d\x), \ \ell_t < \widehat{\ell}|\X_0=\x_0]\\
&=\int_0^{\infty} d\ell \ \gamma\e^{-\gamma\ell}\P[\X_t \in (\x,\x+d\x), \ \ell_t < \ell |\X_0=\x_0]\\
&=\int_0^{\infty} d\ell \ \gamma \e^{-\gamma\ell}\int_0^{\ell} d\ell' [P_0(\x,\ell',t|\x_0)d\x].
\end{align*}
Using the identity
\begin{equation}
\int_0^{\infty}d\ell \ f(\ell)\int_0^{\ell} d\ell' \ g(\ell')=\int_0^{\infty}d\ell' \ g(\ell')\int_{\ell'}^{\infty} d\ell \ f(\ell)
\label{fg}
\end{equation}
for arbitrary integrable functions $f,g$, it follows that
\begin{equation}
\label{bob}
p_0(\x,t|\x_0)=\int_0^{\infty} \e^{-\gamma\ell}P_0(\x,\ell,t|\x_0)d\ell.
\end{equation}
The probability density $p_0(\x,t|\x_0)$ can be expressed in terms of the Laplace transform of the propagator $P_0(\x,\ell,t|\x_0)$ with respect to the local time $\ell$, since the Robin boundary condition maps to an exponential law for the stopping local time $\widehat{\ell}_t$.
The advantage of this formulation is that one can consider a more general probability distribution $\Psi(\ell) = \P[\ellh>\ell]$ for the stopping local time $\ellh$ such that \cite{Grebenkov19b,Grebenkov20,Grebenkov21}
  \begin{equation}
  \label{Boo}
  p_0^{\Psi}(\x,t|\x_0)=\int_0^{\infty} \Psi(\ell)P_0(\x,\ell,t|\x_0)d\ell.
  \end{equation}

Equation (\ref{Boo}) accommodates a wider class of surface reactions. For example, As highlighted in Ref. \cite{Grebenkov20}, one of the possible mechanisms for a non-exponential stopping local time distribution is an encounter-dependent reactivity. This could represent, for example, a progressive activation/deactivation or aging of the reactive surface following each attempted reaction. In order to understand such a process, it is useful to model diffusion as a discrete-time random walk on a hypercubic lattice ${\mathbb Z}^d$ with lattice spacing $a$. First consider a constant reaction rate $\kappa_0$ and introduce the so-called reaction length $\xi=D/\kappa_0$. At a bulk site, a particle jumps to one of the neighboring sites with probability $1/2d$, whereas at a boundary site it either reacts with probability $\Pi=(1+\xi/a)^{-1}\approx a\kappa_0/D$ or return to a neighboring bulk site with probability $1-\Pi$. Assuming the random jumps are independent of the reaction events, the random number of jumps $\widehat{N}$ before a reaction occurs is given by a geometric distribution: $\P[\widehat{N}=n]=\Pi(1-\Pi)^n$, integer $n\geq 0$. In particular, $\E[\widehat{N}]=(1-\Pi)/\Pi=\xi/a$. Introducing the rescaled random variable $\widehat{\ell }=a\widehat{N}$, one finds that \cite{Grebenkov19a}
 \begin{align*}
  \P[\widehat{\ell }\geq \ell]&=\P[\widehat{N}\geq \ell/a]=(1-\Pi)^{\ell/a} =(1+a/\xi)^{-\ell/a} \underset{a\rightarrow 0}\rightarrow \e^{-\ell/\xi}.
  \end{align*}
 That is, for sufficiently small lattice spacing $a$, a reaction occurs (the random walk is terminated) when the random number of realized jumps from boundary sites, multiplied by $a$, exceeds an exponentially distributed random variable (stopping local time) $\widehat{\ell}$ with mean $\xi$. Assuming that a partially reflected random walk on a lattice converges to a well-defined continuous process in the limit $a\rightarrow 0$ (see Refs. \cite{Papanicolaou90,Milshtein95}), one can define partially reflected Brownian motion as reflected Brownian motion stopped at the random time (\ref{Tell0}) \cite{Grebenkov06,Grebenkov07,Grebenkov19a}
where the local time $\ell_t$ is the continuous analog of the rescaled number of surface encounters, $a\widehat{N}$), and $\P[\widehat{\ell}>\ell]=\e^{-\ell/\xi}$. Now suppose that at the $n$th encounter, the reaction probability is $\Pi_n\approx a\kappa_n/D$, with some prescribed sequence of reactivities. Again, assuming that the reaction events are independent,
\begin{align*}
  \P[\widehat{N}=n]=(1-\Pi_0)(1-\Pi_1)\ldots (1-\Pi_{n-1})\Pi_n,\quad n=0,1,2.\ldots
  \end{align*}
Taking the limit $a\rightarrow 0$, we find that $\kappa_n\rightarrow \kappa(\ell)$ and
\begin{equation}
\label{kapell}
\Psi(\ell)=\exp\left (-\frac{1}{D}\int_0^{\ell}\kappa(\ell')d\ell'\right ).
\end{equation}

In the absence of resetting, the propagator $P_0(\x,\ell,t|\x_0)$ satisfies the boundary value problem (BVP) \cite{Grebenkov20,Bressloff22}
\begin{subequations}
\begin{align}
\label{Ploc1}
 &\frac{\partial P_0(\x,\ell,t|\x_0)}{\partial t}=D\nabla^2 P_0(\x,\ell,t|\x_0),\ \x \in \R^d\backslash \calU,\\
\label{Ploc2} &-D\nabla P_0(\x,\ell,t|\x_0) \cdot \n= D\delta(\ell)  P_0(\x,\ell=0,t|\x_0)  +D\frac{\partial}{\partial \ell} P_0(\x,\ell,t|\x_0),\  \x\in \partial \calU,\\
\label{Ploc3}
 & P_0(\x,\ell=0,t|\x_0)=-\nabla p_{0,\infty}(\x,t|\x_0)\cdot \n ,\ \x\in \partial \calU, \end{align}
\end{subequations}
together with the initial condition $P_0(\x,\ell,0|\x_0)=\delta(\x-\x_0)\delta(\ell)
$. Here $p_{0,\infty}$ is the probability density in the case of a totally absorbing target: 
\begin{subequations} 
\begin{align}
\label{popoa}
	&\frac{\partial p_{0,\infty}(\x,t|\x_0)}{\partial t} = D\nabla^2 p_{0,\infty}(\x,t|\x_0), \, \x\in \R^d\backslash \calU,\\
&p_{0,\infty}(\x,t|\x_0)=0,\  \x\in \partial \calU,\ p_{0,\infty}(\x,0|\x_0)=\delta(\x-\x_0).
\label{popob}
	\end{align}
	\end{subequations} 
	The equality (\ref{Ploc3}) can be derived by noting that a constant reactivity is equivalent to a Robin boundary condition, see equation (\ref{bob}). In particular, the Robin boundary condition can be rewritten as
\begin{align}
 \nabla p_0(\x,t|\x_0)\cdot \n&=-\gamma p_0(\x,t|\x_0)=-\gamma \int_0^{\infty}\e^{-\gamma \ell}P_0(\x,\ell,t|\x_0)d\ell,\ \x \in \partial \calU
\end{align}
where $\gamma=\kappa_0/D$.
The result follows from taking the limit $\gamma \rightarrow \infty$ on both sides with $p_0\rightarrow p_{0,\infty}$, and noting that $\lim_{\gamma \rightarrow \infty}\gamma \e^{-\gamma \ell}$ is the Dirac delta function on the positive half-line. One way to interpret the boundary condition (\ref{Ploc2}) is that, for $\ell >0$, the rate at which the local time is increased at a point $\x\in \partial \calU$ is equal to the corresponding flux density at that point, whereas the local time does not change in the bulk of the domain. In addition, the flux when $\ell =0$ is identical to the one for a totally absorbing surface. 

\subsection{Diffusion-mediated surface reactions with position resetting}

Now suppose that the particle can reset to $\x_0\in \R^d\backslash \calU$ at a Poisson rate $r$ when it is in the bulk of the  domain. (Resetting does not occur at the surface boundary.) Denote the corresponding propagator by $P(\x,\ell,t|\x_0)$. Since resetting at a time $\tau$ does not change the accumulation time $\ell_{\tau}$, resetting can be introduced into the BVP for the propagator as follows:
\begin{subequations}
\begin{align}
 &\frac{\partial P(\x,\ell,t|\x_0)}{\partial t}=D\nabla^2 P(\x,\ell,t|\x_0)-rP(\x,\ell,t|\x_0)\nonumber \\
 &\hspace{3cm} +rQ_r(\x_0,\ell,t)\delta(\x-\x_0),\ \x \in \R^d\backslash \calU,
 \label{rPloc1}\\
\label{rPloc2} &-D\nabla P(\x,\ell,t|\x_0) \cdot \n= D \delta(\ell)  P(\x,\ell=0,t|\x_0) +D\frac{\partial}{\partial \ell} P(\x,\ell,t|\x_0),\  \x\in \partial \calU,\\
\label{rPloc3}
 & P(\x,\ell=0,t|\x_0)=-\nabla p_{\infty}(\x,t|\x_0)\cdot \n ,\ \x\in \partial \calU, 
\end{align}
\end{subequations}
where $P(\x,\ell,0|\x_0)=\delta(\x-\x_0)\delta(\ell)$ and $p_{\infty}$ is now the probability density in the case of a totally absorbing target with resetting: 
\begin{subequations} 
\begin{align}
 	&\frac{\partial p_{\infty}(\x,t|\x_0)}{\partial t} = D\nabla^2 p_{\infty}(\x,t|\x_0)-r p_{\infty}(\x,t|\x_0)\nonumber \\
\label{pinf2} &\hspace{3cm}+rQ_{r,\infty}(\x_0,t) \delta(\x-\x_0), \, \x\in \R^d\backslash \calU,\\
 &p_{\infty}(\x,t|\x_0)=0,\  \x\in \partial \calU,\ p_{\infty}(\x,0|\x_0)=\delta(\x-\x_0).
\label{pinf22}
	\end{align}
	\end{subequations} 
	We have also introduced the survival probabilities
	\begin{align}
 Q_r(\x_0,\ell,t)=\int_{\R^d\backslash \calU}P(\x,\ell,t|\x_0)d\x,\quad  Q_{r,\infty}(\x_0,t)=\int_{\R^d\backslash \calU}p_{\infty}(\x,t|\x_0)d\x.
	\end{align}
	It is important to note that the local time is not reset, which means that we cannot use renewal theory to express the propagator with resetting in terms of its counterpart without resetting.

Given the solution to the BVP, the corresponding marginal density for particle position can be obtained from the analog of equation (\ref{oo}), 
 \begin{equation}
  \label{oo}
  p^{\Psi}(\x,t|\x_0)=\int_0^{\infty} \Psi(\ell)P(\x,\ell,t|\x_0)d\ell.
  \end{equation}
  Multiplying equations (\ref{rPloc1}) and (\ref{rPloc2}) by $\Psi(\ell)$ and integrating with respect to $\ell$ gives
\begin{subequations}
\begin{align}
\label{rploc1}
 &\frac{\partial p^{\Psi}(\x,t|\x_0)}{\partial t}=D\nabla^2 p^{\Psi}(\x,t|\x_0)-rp^{\Psi}(\x,t|\x_0)\nonumber \\
 &\hspace{3cm}+rQ_r^{\Psi}(\x_0,t)\delta(\x-\x_0),\ \x \in \R^d\backslash \calU,
 \\
\label{rploc2} &-D\nabla p^{\Psi}(\x,t|\x_0) \cdot \n= D\int_0^{\infty}\psi(\ell) P(\x,\ell,t|\x_0)d\ell,\  \x\in \partial \calU,
\end{align}
\end{subequations}
with
\begin{equation}
\label{pP}
\psi(\ell)=-\frac{d\Psi(\ell)}{d\ell},\quad \widetilde{\psi}(q)=1-q\widetilde{\Psi}(q)
\end{equation}
and
\begin{equation}
\label{Qpsi}
Q^{\Psi}_r(\x_0,t)=\int_0^{\infty}\Psi(\ell)Q_r(\x_0,\ell,t)d\ell=\int_{\R^d\backslash \calU}p^{\Psi}(\x,t|\x_0)d\x.
\end{equation}
We have used integration by parts and the identity $\Psi(0)=1$. It is clear that equations (\ref{rploc1}) and (\ref{rploc2}) do not form a closed BVP for $p^{\Psi}$. The only exception is the exponential case, since $\psi(\ell)=\gamma \Psi(\ell)$. Integrating equation (\ref{rploc2}) with respect to points on the boundary, determines a relation between the total flux into the surface and the propagator:
\begin{equation}
\label{JP}
 {J}^{\Psi}_r(\x_0,t)\equiv -D\int_{\partial \calU} \nabla p^{\Psi}(\x,t|\x_0)\cdot \n\, d\sigma=D\int_0^{\infty}\psi(\ell)\left [ \int_{\partial \calU}P(\x,\ell,t|\x_0)d\sigma \right ]d\ell.
\end{equation}
Similarly, integrating equation (\ref{rploc1}) with respect to $\x\in \R^d\backslash \calU$ yields a relation between the surface flux and the survival probability,
\begin{align}
\label{JQ}
\frac{\partial Q^{\Psi}(\x_0,t)}{\partial t}=-{J}^{\Psi}_r(\x_0,t).
\end{align}

As in the analysis section 2, it will be more convenient to work in Laplace space. Laplace transforming equations (\ref{rPloc1})--(\ref{rPloc3}) gives
\begin{subequations}
\begin{align}
\label{rPloc1LT}
 & D\nabla^2 \PP(\x,\ell,s|\x_0)-(r+s)\PP(\x,\ell,s|\x_0)\\
 &\quad =-[\delta(\ell)+r\Q_r(\x_0,\ell,s)]\delta(\x-\x_0),\ \x \in \R^d\backslash \calU,\nonumber \\
\label{rPloc2LT} &-D\nabla \PP(\x,\ell,s|\x_0) \cdot \n= D \PP(\x,\ell=0,s|\x_0) \ \delta(\ell)  +D\frac{\partial}{\partial \ell} \PP(\x,\ell,s|\x_0),\  \x\in \partial \calU,\\
\label{rPloc3LT}
 & \PP(\x,\ell=0,s|\x_0)=-\nabla \p_{\infty}(\x,s|\x_0)\cdot \n ,\ \x\in \partial \calU.
\end{align}
\end{subequations}
Furthermore, Laplace transforming equations (\ref{pinf2}) and (\ref{pinf22}) shows that $\p_{\infty}$ can be expressed as
\begin{equation}
\p_{\infty}(\x,s|\x_0)=[1+r\Q_{r,\infty}(\x_0,s)]G(\x,\alpha|\x_0),\quad \alpha=\sqrt{\frac{r+s}{D}},
\end{equation}
where $G$ is a modified Helmholtz Green's function:
\begin{align}
\label{GGa}
	&D\nabla^2 G(\x,\alpha|\x_0)-(r +s)G(\x,\alpha|\x_0)=-\delta(\x-\x_0), \, \x\in \R^d\backslash \calU,\nonumber \\
 &G(\x,\alpha|\x_0)=0,\  \x\in \partial \calU.
	\end{align}
	(In fact, $\p_{0,\infty}(\x,s|\x_0)=G(\x,\sqrt{s/D}|\x_0)$.)
Next, Laplace transforming equations (\ref{rploc1}) and (\ref{JP}) yields
\begin{subequations}
\begin{align}
\label{rploc1LT}
 &D\nabla^2 \p^{\Psi}(\x,s|\x_0)-(r+s)\p^{\Psi}(\x,s|\x_0)=-[1+r\Q_r^{\Psi}(\x_0,s)]\delta(\x-\x_0),\ \x \in \R^d\backslash \calU,\nonumber \\
 &\\
\label{rploc2LT} &-D\nabla \p^{\Psi}(\x,s|\x_0) \cdot \n= D\int_0^{\infty}\psi(\ell) \PP(\x,\ell,s|\x_0)d\ell,\  \x\in \partial \calU.
\end{align}
\end{subequations}
The corresponding equations without resetting ($r=0$) are
\begin{subequations}
\begin{align}
\label{rploc1LT0}
 &D\nabla^2 \p_0^{\Psi}(\x,s|\x_0)-s\p_0^{\Psi}(\x,s|\x_0)=-\delta(\x-\x_0),\ \x \in \R^d\backslash \calU,  \\
 & 
-D\nabla \p_0^{\Psi}(\x,s|\x_0) \cdot \n= D\int_0^{\infty}\psi(\ell) \PP_0(\x,\ell,s|\x_0)d\ell,\  \x\in \partial \calU.
\label{rploc2LT0} 
\end{align}
\end{subequations}
Comparing equations (\ref{rploc1LT}) and (\ref{rploc1LT0}) implies that 
\begin{equation}
\label{urr}
\p^{\Psi}(\x,s|\x_0)=[1+r\Q_r^{\Psi}(\x_0,s)]\p_0^{\Psi}(\x,r+s|\x_0)+u^{\Psi}_r(\x,s|\x_0),
\end{equation}
where $u^{\Psi}_r$ satisfies the homogeneous version of equation (\ref{rploc1LT}) together with a non-trivial boundary condition on $\partial \calU$. 

In the special case of a constant reactivity (Robin boundary condition), the term $u^{\Psi}_r$ is identically zero. In order to show this, note that the propagator $P$ satisfies the integral equation
\begin{align}
 P(\x,\ell,t|\x_0)&=\e^{-rt}P_0(\x,\ell,t|\x_0)\nonumber \\
 &\quad +r\int_0^{\ell} \left (\int_0^t \e^{-r\tau}P_0(\x,\ell-\ell',\tau|\x_0)Q_r(\x_0,\ell',t-\tau)d\tau \right )d\ell'.
\end{align}
The first term on the right-hand side represents all trajectories that do not reset in the interval $[0,t]$. The double integral represents the complementary set of trajectories that reset to $\x_0$ at least once. It is assumed that the last reset occurs at time $t-\tau$ with the particle having spent an amount of time $\ell'$ in a neighborhood of the boundary without being absorbed. This occurs with probability density $Q_r(\x_0,\ell',t-\tau)$. Over the time interval $[t-\tau,t]$ there are no more resettings and the local time increases by an additional amount $\ell-\ell'$ with associated probability density $P_0(\x,\ell-\ell',\tau|\x_0)$. Laplace transforming the integral equation using the convolution theorem shows that
\begin{align}
\label{okk}
 \PP(\x,\ell,s|\x_0)=\PP_0(\x,\ell,r+s|\x_0)+r\int_0^{\ell} \PP_0(\x,\ell-\ell',r+s|\x_0)Q_r(\x_0,\ell',s)d\ell' .
\end{align}
{Let us perform a second Laplace transform with respect to the local time $\ell$ by setting
\begin{subequations}
\begin{align}
 \calP(\x,z,s|\x_0)&=\int_0^{\infty}\e^{-z\ell}\PP(\x,\ell,s|\x_0)d\ell,\\
  \calQ_r(\x_0,z,s)&=\int_0^{\infty}\e^{-z\ell}\Q_r(\x_0,\ell,s)d\ell=\int_{\R^d\backslash \calU}\calP(\x,z,s|\x_0)d\x.
 \end{align}
 \end{subequations}
Multiplying both sides of equation (\ref{okk}) by $\e^{-z \ell}$ and applying the convolution theorem to the $\ell$-Laplace transform implies that
\begin{equation}
\label{calpspec}
\calP(\x,z,s|\x_0)=[1+r\calQ_r(\x_0,z,s)]\calP_0 (\x,z,r+s|\x_0).
\end{equation}
Integrating both sides of this equation with respect to $\x\in \Omega\backslash \calU$ gives
\begin{align}
\label{Qr0gen}
   \calQ_r(\x_0, z,s) =  [1+r\calQ_r(\x_0,z,s)]\calQ_0(\x_0,z,r+s),
\end{align} 
which can be rearranged to show that
\begin{equation}
\label{Qgen}
\calQ_r(\x_0,z,s)=\frac{\calQ_0(\x_0,z,r+s)}{1-r\calQ_0(\x_0,z,r+s)}.
\end{equation}
}

{We now make the following observations. If we make the identification $z=\kappa_0$, where $\kappa_0$ is the constant reactivity for the Robin boundary condition, then $\calQ_r(\x_0,\kappa_0,s)=\Q_r^{\Psi}(\x_0,s)$ and
\begin{align}
\Q_r^{\Psi}(\x_0 ,s) &=\frac{\displaystyle \Q_0^{\Psi}(\x_0, r+s)}{1-r\Q_0^{\Psi}(\x_0, r+s)},\quad \Psi=\e^{-\kappa_0\ell/D}.
\label{Qfin}
\end{align}
Comparison with equation (\ref{urr}) establishes that the homogeneous solution $u_r^{\Psi}=0$.
On the other hand, for a non-exponential distribution $\Psi(\ell)$, it is first necessary to invert equation (\ref{Qgen}) to determine $\Q_r(\x_0,\ell,s)$ and then use this to calculate $\Q_r^{\Psi}(\x_0 ,s) $. Clearly, equation (\ref{Qfin}) no longer holds and hence $u_r^{\Psi}\neq 0$. The inverse Laplace transform is given by a Bromwich integral of the form
\begin{align}
\Q_r(\x_0,\ell,s)&=\frac{1}{2\pi i}\int_{c-i\infty}^{c+i\infty}\e^{z\ell}\calQ_r(\x_0,z,r)dz\nonumber \\
&=\frac{1}{2\pi i}\int_{c-i\infty}^{c+i\infty}\e^{z\ell}\frac{\calQ_0(\x_0,z,r+s)}{1-r\calQ_0(\x_0,z,r+s)}dz.
\end{align}
The real constant $c$ is chosen so that the Bromwich contour lies to the right of all poles in the complex $z$-plane. One can then close the contour to the right and express $\Q_r(\x_0,\ell,s)$ in terms of the sum of the residues arising from the poles of the function $1-r\Q_0(\x_0,z,r+s)$ in the $z$-plane. The latter is itself obtained by solving the BVP for the diffusion equation with a Robin boundary condition on $\partial \calU$ and no resetting, see section 2.
}

\subsection{Diffusion-mediated surface reactions with position and local time resetting}

One situation where equation (\ref{Qfin}) holds for a general distribution $\Psi$ is if the local time is also reset so that $(\X_t,\ell_t)\rightarrow (\x_0,0)$ at a Poisson rate $r$ prior to absorption. This does not mean resetting the number of encounters between particle and boundary, since such a quantity is accumulative. However, suppose that there is some internal state of the particle that is modified whenever it is in a neighborhood of $\partial \calU$, and that this modification is proportional to the local time. Moreover, assume that the reactivity depends on the current internal state. Resetting the internal state to its initial value whenever the particle resets to $\x_0$ is equivalent to resetting the reactivity and thus the effective local time.
 Incorporating position and local time resetting into the BVP given by equations (\ref{Ploc1})--(\ref{Ploc3}), we have
\begin{subequations}
\begin{align}
 &\frac{\partial P_r(\x,\ell,t|\x_0)}{\partial t}=D\nabla^2 P_r(\x,\ell,t|\x_0)-rP_r(\x,\ell,t|\x_0)\nonumber \\
 &\hspace{3cm} +r\delta(\x-\x_0)\delta(\ell),\ \x \in \Omega\backslash \calU, \label{rPloc1n} \\
 &-D\nabla P_r(\x,\ell,t|\x_0) \cdot \n= D P_r(\x,\ell=0,t|\x_0) \ \delta(\ell)  +D\frac{\partial}{\partial \ell} P_r(\x,\ell,t|\x_0) \mbox{ for }  \x\in \partial \calU,\nonumber \\
 \label{rPloc3n}
\end{align}
\end{subequations}
The unknown $P_r(\x,\ell=0,t|\x_0) $ for $\x \in \partial \calU$ is determined by noting that the
stochastic resetting process is memoryless, and hence the propagator satisfies a first renewal equation of the form
\begin{equation}
 P_r(\x,\ell,t|\x_0)= \e^{-rt}P_0(\x,\ell,t|\x_0)+r\int_0^t\e^{-r \tau}P_r(\x,\ell,t-\tau|\x_0)d\tau .
\end{equation}
The first term on the right-hand side represents all trajectories that do not undergo any resettings, which occurs with probability $\e^{-rt}$. The second term represents the complementary set of trajectories that reset at least once with the first reset occurring at time $\tau$. Laplace transforming the renewal equation and rearranging shows that
\begin{equation}
\label{con}
\PP_r(\x,\ell,s|\x_0)=\left (1+\frac{r}{s}\right )\PP_0(\x,\ell,r+s|\x_0).
\end{equation}
Since $\PP_0(\x,\ell=0,s|\x_0)=-\nabla p_{\infty}(\x,s|\x_0) \cdot \n$ for $\x\in \calU$, it follows that
\begin{equation}
\PP_0(\x,\ell=0,s|\x_0)=-\left (1+\frac{r}{s}\right )\nabla p_{\infty}(\x,s|\x_0) \cdot \n,\quad \x \in \partial \calU.
\end{equation}
Multiplying both sides of equation (\ref{con}) by $s$ and taking the limit $s\rightarrow 0$, then establishes that there exists a non-equilibrium stationary state (NESS) $P_r^*(\x,\ell|\x_0)$:
\begin{equation}
\label{con0}
 P_r^*(\x,\ell|\x_0)=\lim_{t\rightarrow \infty}P_r(\x,\ell,t|\x_0)=\lim_{s\rightarrow 0}s \PP_r(\x,\ell,s|\x_0)=r\PP_0(\x,\ell,r|\x_0).
\end{equation}

In contrast to position resetting alone, one cannot simply take
$p_r^{\Psi}(\x,t|\x_0)=\int_0^{\infty}\Psi(\ell)P_r(\x,\ell,t|\x_0)d\ell$, since $\ell_t$ is no longer a monotonically increasing function of time $t$. Therefore, we proceed by partitioning the set of contributing paths according to the number of resettings and for a given number of resettings decomposing the path into time intervals over which $\ell_t$ is monotonically increasing. Let ${\mathcal I}_t$ denote the number of resettings in the interval $[0,t]$ and let ${\mathcal T}=\inf\{t>0, \ell_t >\widehat{\ell}\}$. Then
\begin{align}
 p_r^{\Psi}(\x,t|\x_0)d\x&=\e^{-rt}\P[\X_t \in [\x,\x+d\x]|\X_0=\x_0,\, {\mathcal T}>t,\, {\mathcal I}_t=0]\\
 &\quad +r\e^{-rt}\P[\X_t \in [\x,\x+d\x]|\X_0=\x_0,\, {\mathcal T}>t,\, {\mathcal I}_t=1]\nonumber \\
 &\quad +r^2\e^{-rt}\P[\X_t \in [\x,\x+d\x]|\X_0=\x_0,\, {\mathcal T}>t,\, {\mathcal I}_t=2]+\ldots \nonumber 
\end{align}
That is,
\begin{align}
 p_r^{\Psi}(\x,t|\x_0)&=\e^{-rt} p_0^{\Psi}(\x,t|\x_0)+r\e^{-rt} \int_0^t p_0^{\Psi}(\x,\tau|\x_0)Q_0^{\Psi}(\x_0,t-\tau)d\tau\\
 &\quad +r\e^{-rt} \int_0^t \int_0^{t-\tau}p_0^{\Psi}(\x,\tau|x_0)Q_0^{\Psi}(\x_0,t-\tau)Q_0^{\Psi}(\x_0,t-\tau-\tau')d\tau'd\tau+\ldots \nonumber
\end{align}
where $Q_0^{\Psi}$ is the survival probability without resetting. Laplace transforming the above equation and using the convolution theorem shows that
\begin{align}
 \p_r^{\Psi}(\x,s|\x_0)&=\p_0^{\Psi}(\x,r+s|\x_0)+r\p_0^{\Psi}(\x,r+s|\x_0)\Q_0^{\Psi}(x_0,r+s)\nonumber \\
&\quad +r^2\p_0(\x,r+s|\x_0)\Q_0(x_0,r+s)^2+\ldots
\end{align}
Summing the geometric series thus yields the result
\begin{equation}
\label{prQ}
\p_r^{\Psi}(\x,s|\x_0)=\frac{\p_0^{\Psi}(\x,r+s|\x_0)}{1-r\Q^{\Psi}_0(\x_0,r+s)}.
\end{equation}
Finally, integrating both sides with respect to $\x\in \Omega \backslash \calU$ yields equation (\ref{Qfin}) for general $\Psi$.

\setcounter{equation}{0}

\section{A partially reactive spherical surface}

Let us return to the example of a spherical surface in $\R^d$, which was analyzed in section 2.2 in the case of Robin boundary conditions and an initial condition distributed uniformly on the sphere of radius $\rho_0$. We begin by considering the propagator BVP for position resetting.
Laplace transforming equations (\ref{rPloc1})-(\ref{rPloc3}) and introducing spherical polar coordinates gives
  \begin{subequations}
\begin{align}
\label{rspha}
   &D\frac{\partial^2\PP}{\partial \rho^2} + D\frac{d - 1}{\rho}\frac{\partial \PP}{\partial \rho} -(r+s)\PP(\rho,\ell, s|\rho_0) \nonumber \\ 
 & \hspace{3cm}=-[\delta(\ell)+r\Q_r(\rho_0,\ell,s) ]\Gamma_d\delta(\rho-\rho_0) ,\quad \rho_1<\rho,\\
 &\frac{\partial }{\partial \rho}\PP(\rho,\ell,s|\rho_0) =  \PP(\rho,\ell=0,s|\rho_0) \ \delta(\ell)  +\frac{\partial}{\partial \ell} \PP(\rho,\ell,s|\rho_0),\  \rho=\rho_1,
 \label{rsphb}\\
  & \PP(\rho_1,\ell=0,s|\rho_0)=\left . \frac{d}{d\rho}\p_{\infty}(\rho,s|\rho_0) \right |_{\rho=\rho_1} ,
 \label{rsphc}
\end{align}
\end{subequations}
where
\begin{align}
 \Q_r(\rho_0,\ell,s)=\Omega_d \int_{\rho_1}^{\infty}\rho^{d-1}\PP(\rho,\ell,s|\x_0)d\rho.
	\end{align}
The corresponding density $\p_{\infty}(\rho,s|\rho_0)$ for a totally absorbing surface is given by equations (\ref{pr}) and (\ref{pinf}):
\begin{equation}
\p_{\infty}(\rho,s|\rho_0)= [1+r\Q_{r,\infty}(\rho_0,s)]G(\rho, \alpha| \rho_0),
\end{equation}
where $G$ is the Green's function defined by equations (\ref{Ga}) and (\ref{Gb}). The general solution of
equations (\ref{rspha})--(\ref{rsphc}) can be written in the form
  \begin{align}
\label{rqir}
    \PP(\rho,\ell, s|\rho_0) = C_r(\ell,s)\rho^\nu K_\nu(\alpha \rho) +  \left [\delta(\ell)+r\Q_r(\rho_0,\ell,s)\right ]G(\rho, \alpha| \rho_0) , \ \rho_1\leq \rho.
\end{align}
The unknown coefficient $C_r(\ell,s)$ is determined from the boundary condition (\ref{rsphb}) and (\ref{rsphc}):
\begin{align}
\label{C}
 & \frac{dC_r(\ell,s)}{d\ell} F_{\alpha}(\rho_1) = C_r(\ell,s) F_{\alpha}'(\rho_1)\\
  &\quad +  r\left [\Q_r(\rho_0,\ell,s)-\delta(\ell)\Q_{r,\infty}(\rho_0,s)\right ] \left . \frac{d}{d\rho}G(\rho,\alpha|\rho_0) \right |_{\rho=\rho_1} ,\nonumber
\end{align}
with $F_{\alpha}(\rho)$ given by equation (\ref{FF}).
Equation (\ref{C}) has the solution
\begin{align}
\label{cell0}
  C_r(\ell,s)&=  C_r(0,s)  \e^{-\Lambda(\alpha)\ell}\\
 &\quad + r\chi(\alpha)\int_0^{\ell}\e^{-\Lambda(\alpha)(\ell-\ell')}\left [\Q_r(\rho_0,\ell',s)-\delta(\ell')\Q_{r,\infty}(\rho_0,s)\right ] d\ell',\nonumber 
\end{align}
where
\begin{equation}
\label{tilSa}
\chi(\alpha)=\frac{1}{F_{\alpha}(\rho_1)} \left . \frac{d}{d\rho}G(\rho,\alpha|\rho_0) \right |_{\rho=\rho_1} ,
\end{equation}
and
\begin{equation}
\label{tilSb}
\Lambda(\alpha)=-\frac{F_{\alpha}'(\rho_1)}{F_{\alpha}(\rho_1)}=-\frac{\nu}{\rho_1}-\frac{\alpha  K'_{\nu}(\alpha \rho_1)}{ K_{\nu}(\alpha \rho_1)}.
\end{equation}
In addition, setting $\ell=0$ and $\rho=\rho_1$ in equation (\ref{rqir}) and using (\ref{rsphc}) implies that
\begin{equation}
C_r(0,s)=[1+r \Q_{r,\infty}(\rho_0,s)]\chi(\alpha),
\end{equation}
and, hence,
\begin{align}
\label{cell}
 C_r(\ell,s)&=\chi(\alpha)\left \{ \e^{-\Lambda(\alpha)\ell}+r\int_0^{\ell}\e^{-\Lambda(\alpha)(\ell-\ell')}\Q_r(\rho_0,\ell',s)d\ell'\right \}.
\end{align}

Substituting for $C_r(\ell,s)$ into equation (\ref{rqir}) and integrating over the domain $\Omega\backslash \calU$ yields an integral equation for $\Q_r(\rho_0,\ell,s)$:
\begin{align}
    \Q_r(\rho_0,\ell,s) &=\overline{F}_{\alpha}  \chi(\alpha)\left \{ \e^{-\Lambda(\alpha)\ell}+r\int_0^{\ell}\e^{-\Lambda(\alpha)(\ell-\ell')}\Q_r(\rho_0,\ell',s)d\ell'\right \} \nonumber \\
 \label{rqir3}
  &\quad +  \left [\delta(\ell)+r\Q_r(\rho_0,\ell,s)d\ell\right ]\overline{G}_{\alpha}(\rho_0) ,
 \end{align}
where
\begin{equation}
\overline{F}_{\alpha}(\rho_1)=\Omega_d\int_{\rho_1}^{\infty}\rho^{d-1}F_{\alpha}(\rho)d\rho,
\end{equation}
and
\begin{equation}
\overline{G}_{\alpha}(\rho_0)=\Omega_d\int_{\rho_1}^{\infty}\rho^{d-1}G(\rho,\alpha|\rho_0) d\rho.
\end{equation}
It follows that
\begin{align}
  \Q_0(\rho_0,\ell,r+s) &=\overline{F}_{\alpha}  \chi(\alpha)  \e^{-\Lambda(\alpha)\ell}+   \delta(\ell) \overline{G}_{\alpha}(\rho_0) .
 \end{align}
 Laplace transforming (\ref{rqir3}) with respect to $\ell$ yields the analog of equation (\ref{Qgen}), namely,
 \begin{equation}
\label{Qgen2}
\calQ_r(\rho_0,z,s)=\frac{\calQ_0(\rho_0,z,r+s)}{1-r\calQ_0(\rho_0,z,r+s)}
\end{equation}
with
\begin{equation}
\calQ_0(\rho_0,z,r+s)=\frac{\overline{F}_{\alpha}(\rho_1)  \chi(\alpha)}{z+\Lambda(\alpha)}  + \overline{G}_{\alpha_s}(\rho_0)
\end{equation}

We could now proceed to determine $ \Q_r(\rho_0,\ell,s)$ by inverting the $z$-transform on the right-hand side of equation (\ref{Qgen2}), and then calculating the associated survival probability $ \Q_r^{\Psi}(\rho_0,s)$. Here we will consider the simpler case of position and local time resetting for which 
 \begin{equation}
\label{Qgen3}
\Q_r^{\Psi}(\rho_0,s)=\frac{\Q_0^{\Psi}(\rho_0,r+s)}{1-r\Q_0^{\Psi}(\rho_0,r+s)},
\end{equation}
with
\begin{align}
\Q_0^{\Psi}(\rho_0,r+s)&=\int_0^{\infty}\Psi(\ell) \left \{\overline{F}_{\alpha}(\rho_1)   \chi(\alpha)  \e^{-\Lambda(\alpha)\ell}+   \delta(\ell) \overline{G}_{\alpha}(\rho_0) \right \}\nonumber \\
&= \overline{G}_{\alpha}+ \overline{F}_{\alpha}(\rho_1)   \chi(\alpha)\widetilde{\Psi}(\Lambda(\alpha)) .
\end{align}
An alternative expression for $\Q_0^{\Psi}$ can be obtained by noting that for $r=0$, 
\begin{equation}
(r+s)\Q_0^{\Psi}(\rho_0,r+s)-1=- \widetilde{J}_0^{\Psi}(\rho_0,r+s),
\end{equation}
and the Laplace-transformed flux into the spherical target is
\begin{align}
\label{zoboe2} 
  \widetilde{J}_0^{\Psi}(\rho_0,r+s) &=  j(\rho_0,\alpha) \widetilde{\psi}(\Lambda(\alpha)).
  \end{align}
where, see equations (\ref{boll}) and (\ref{Jinf}),
\begin{equation}
\label{jj}
j(\rho_0,\alpha)\equiv D\Omega_d\rho_1^{d-1} \left . \frac{d}{d\rho}G(\rho,\alpha|\rho_0) \right |_{\rho=\rho_1} =\frac{\rho_0^{\nu}K_{\nu}(\alpha \rho_0)}{\rho_1^{\nu}K_{\nu}(\alpha \rho_1)}.
\end{equation}

Having obtained $Q_r^{\Psi}(\rho_0 ,s)$, the MFPT (if it exists) can be obtained by setting $s=0$ in equation (\ref{Qfin}). Since the domain $\R^d\backslash \calU$ is unbounded, the MFPT without resetting is infinite. On the other hand, for $r>0$ we have
\begin{align}
 T_r(\rho_0)&=\frac{\displaystyle \Q_0^{\Psi}(\rho_0, r)}{1-rQ_0^{\Psi}(\rho_0, r)}=\frac{1-  \widetilde{J}_0^{\Psi}(\rho_0,r)}{r  \widetilde{J}_0^{\Psi}(\rho_0,r)}=\frac{1-j(\rho_0,\alpha_r) \widetilde{\psi}(\Lambda(\alpha_r))}{rj(\rho_0,\alpha_r) \widetilde{\psi}(\Lambda(\alpha_r))}
\end{align}
for $\alpha_r=\sqrt{{r}/{D}}$. 
Since $\nu=1/2,0,-1/2$ for $d=1,2,3$, respectively, and $K_{\pm}(z)=\e^{-z}\sqrt{\pi/2z}$, it follows from equations (\ref{tilSb}) and (\ref{jj}) that
\begin{subequations}
\begin{align}
\Lambda(\alpha)= \alpha,\quad &j(\rho_0,\alpha) =\e^{-\alpha(\rho_0-\rho_1)}\quad  \mbox{ for } d=1, \\
\Lambda(\alpha)   =  \frac{\alpha K_{0}'(\alpha\rho_1)}{K_0(\alpha \rho_1)},\quad & j(\rho_0,\alpha)= \frac{K_{0}(\alpha\rho_0)}{K_0(\alpha \rho_1)}\quad  \mbox{ for } d=2,\\
\Lambda(\alpha)= \frac{1}{\rho_1}  +\alpha,\quad & j(\rho_0,\alpha)=\frac{\rho_1}{\rho_0} \e^{-\alpha(\rho_0-\rho_1)}\quad  \mbox{ for } d=3.
\end{align}
\end{subequations}
It remains to specify the stopping local time distribution $\Psi$ and its Laplace transform. For the sake of illustration, we will consider the gamma distribution $\psi_{\rm gam}$ and the Pareto-II (Lomax) distribution $\psi_{\rm par}$:
\begin{equation}
\label{GamD}
\psi_{\rm gam}(\ell)=\frac{\gamma(\gamma \ell)^{\mu-1}\e^{-\gamma \ell}}{\Gamma(\mu)},\ \psi_{\rm par}(\ell)=\frac{\gamma \mu}{(1+\gamma \ell)^{1+\mu}} ,\quad \mu >0,
\end{equation}
where $\gamma=\kappa_0/D$, $\kappa_0$ is some reference reactivity, and $\Gamma(\mu)$ is the gamma function, $\Gamma(\mu)=\int_0^{\infty}\e^{-t}t^{\mu-1}dt$. (Plots of these and other stopping local time densities can be found in \cite{Grebenkov20}.) Note that if $\mu=1$ then
\begin{equation}
\psi_{\rm gam}(\ell)=\gamma \e^{-\gamma \ell},\quad \kappa(\ell) =\kappa_0,
\end{equation}
which recovers the exponential distribution (constant reactivity).
The corresponding Laplace transforms are
\begin{equation}
\widetilde{\psi}_{\rm gam}(q)=\left (\frac{\gamma}{\gamma+q}\right )^{\mu},\quad \widetilde{\psi}_{\rm gam}'(q)=-\mu \left (\frac{\gamma}{\gamma+q}\right )^{\mu}\frac{1}{\gamma+q}.
\end{equation}
and
\begin{subequations}
\begin{align}
 \widetilde{\psi}_{\rm par}(q)&=\mu\left (\frac{q}{\gamma}\right )^{\mu}\e^{q/\gamma}\Gamma(-\mu,q/\gamma),\\
 \widetilde{\psi}_{\rm par}'(q)&=\mu\left (\frac{q}{\gamma}\right )^{\mu}\e^{q/\gamma}\left (\left [\frac{\mu}{q}+\frac{1}{\gamma}\right ]\Gamma(-\mu,q/\gamma)+\partial_{q} \Gamma(-\mu,q/\gamma)\right ).
\end{align}
\end{subequations} 
Here $\Gamma(\mu,z)$ is the upper incomplete gamma function,
\begin{equation}
\quad \Gamma(\mu,z)=\int_z^{\infty}\e^{-t}t^{\mu-1}dt,\ \mu >0.
\end{equation}
It can be seen that $\widetilde{\psi}_{\rm gam}(0)=1$ and $\widetilde{\psi}_{\rm gam}'(0)=-\mu/\gamma<-\infty$. On the other hand, using the identity
\begin{equation}
\Gamma(1-\mu,z)=-\mu \Gamma(-\mu,z) +z^{-\mu}\e^{-z},
\end{equation}
it can be checked that $\widetilde{\psi}(0)=1$, whereas $\widetilde{\psi}'(0)$ is only finite if $\mu>1$. In the latter case
\begin{equation}
-\widetilde{\psi}'(0)=\E[\ell]= \frac{\Gamma(\mu-1)\Gamma(2)}{\gamma \Gamma(\mu)}=\frac{1}{\gamma (\mu-1)}.
\end{equation}
The blow up of the moments when $\mu<1$ reflects the fact that the Pareto-II distribution has a long tail.

\begin{figure}[b!]
\raggedleft
\includegraphics[width=12cm]{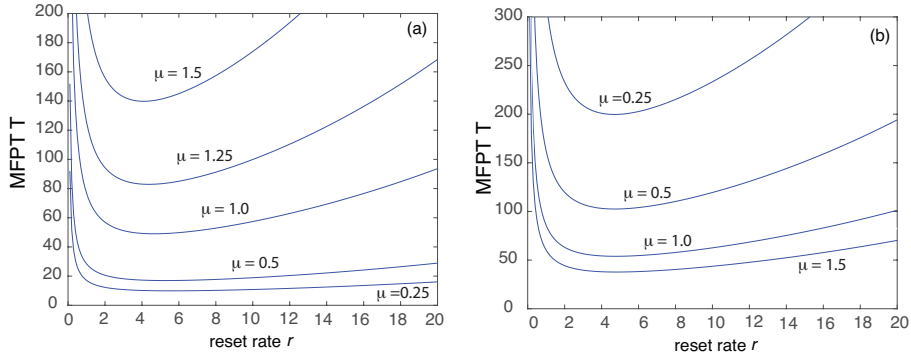} 
\caption{Plots of the MFPT $T_r(\rho_0)$ as a function of the resetting rate $r$ for $d=3$, and various values of the index $\mu$. (a) Gamma distribution. (b) Pareto-II model. We also set $\gamma=\kappa_0/D=1$, $\rho_1=0.2$ and $\rho_0=1$.}
\label{fig4}
\end{figure}

\begin{figure}[t!]
\raggedleft
\includegraphics[width=12cm]{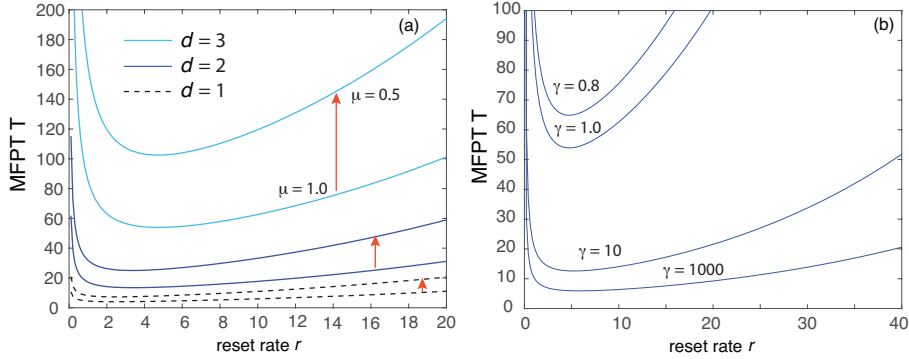} 
\caption{Plots of the MFPT $T_r(\rho_0)$ as a function of the resetting rate $r$ for the Pareto-II model with $\mu=1$. (b) Shift in $T_r(\rho_0)$  in response to a decrease in $\mu$ for $d=1,2,3$. (b) dependence on the reaction rate parameter $\gamma$ for $d=3$. Other parameters are the same as Fig. \ref{fig4}.}
\label{fig5}
\end{figure}

\begin{figure}[t!]
\raggedleft
\includegraphics[width=12cm]{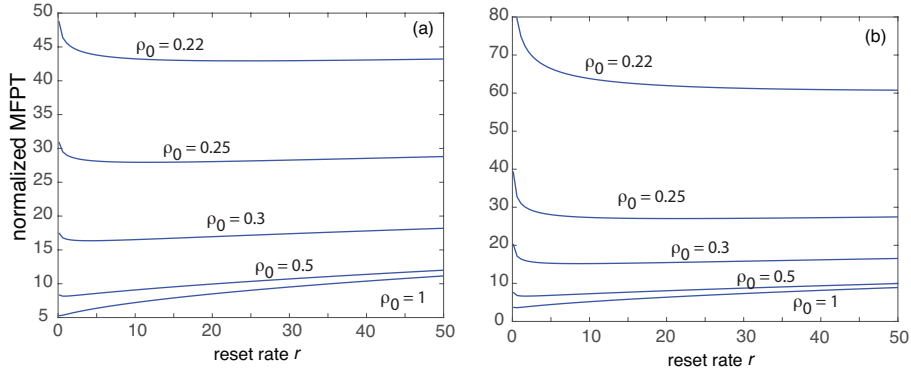} 
\caption{Plots of the normalized MFPT $\Delta T_r(\rho_0)$  as a function of the resetting rate $r$ for the Pareto-II model with $\mu=1$ and various initial radii $\rho_0$ . (a) $d=2$. (b) $d=1$. Other parameters are the same as Fig. \ref{fig4}.}
\label{fig6}
\end{figure}

In Fig. \ref{fig4} we show example plots of $T_r(\rho_0)$ as a function of the resetting rate in the case of the gamma and Pareto-II distributions for fixed $\rho_0=1$ and $\gamma=1$. As expected, the MFPT is a unimodal function of $r$ with a minimum at some optimal value $r_{\rm opt}$ that depends on $\mu$.
The dependence on the spatial dimension is illustrated in Fig. \ref{fig5}(a), where we plot the MFPT for two values of $\mu$ in the cases $d=1,2,3$. It can be seen that the the sensitivity of the MFPT curves to changes in $\mu$ increases significantly with the dimension. The dependence on the reaction rate is illustrated in Fig. \ref{fig5}(b). As expected, reducing $\gamma$ shifts the MFPT resetting curves upwards. In addition, the MFPT approaches zero as $\rho_0\rightarrow \rho_1$ in the large-$\gamma$ limit since the surface becomes totally absorbing. The MFPT of the latter is obtained by setting $\widetilde{\psi}(q)=1$ for all $q$:
\begin{align}
T_{r,\infty}(\rho_0)&=\frac{1-j(\rho_0,\alpha_r) }{rj(\rho_0,\alpha_r)}.
\end{align}
It follows that
\begin{align}
\Delta T_r(\rho_0)&\equiv\frac{T_{r}(\rho_0)}{T_{r,\infty}(\rho_0)}=\frac{1-j(\rho_0,\alpha_r) \widetilde{\psi}(\Lambda(\alpha_r))}{[1-j(\rho_0,\alpha_r) ]\widetilde{\psi}(\Lambda(\alpha_r))}.
\end{align}
Note that $\Delta T_r(\rho_0)\rightarrow 1$ in the limit $\gamma \rightarrow \infty$. Example plots of the normalized MFPT $\Delta T_r(\rho_0)$ as a function of $r$ are shown in Fig. \ref{fig6}. It can be seen that as $\rho_0$ approaches $\rho_1$, the MFPT resetting curve switches from a monotonically increasing function of $r$ to a unimodal function with a minimum at some nonzero value of $r$.

\section{Discussion}  

In this paper we have shown how combining stochastic resetting with generalized diffusion-mediated surface reactions leads to a non-trivial boundary value problem for the joint probability density or generalized propagator of the particle position and the boundary local time. If only the position of the particle is reset, then resetting is not governed by a renewal process, and one has to take a double transform with respect to $t$ and the local time in order to solve the propagator BVP. It is then necessary to invert the Laplace transform with respect to the local time in order to determine the corresponding survival probability. On the other, simultaneous position and local time resetting is governed by a renewal process and one can express the survival probability with resetting in terms of the survival probability without resetting. The analysis of the MFPT for absorption is then relatively straightforward. We illustrated the latter case using the example of a spherically symmetric surface. In addition to determining the optimal resetting rate that minimizes the MFPT, we also showed that the relative increase in the MFPT compared to the case of a totally absorbing surface can itself exhibit non-monotonic variation with $r$. In future work, we will explore how these results are modified when only the position of the particle is reset. Another issue is to identify possible physical mechanisms for local time resetting, based on the hypothesis that the rate of absorption depends on some internal state of the particle.

As we have recently shown elsewhere \cite{Bressloff22}, it is also possible to develop a theory of diffusion-mediated reactions in the case of targets whose interiors are partially absorbing. Now the particle can freely enter and exit $\calU$, and is absorbed probabilistically when inside $\calU$. The main difference between absorption by the target boundary and target interior is that the latter involves the occupation time (accumulated time that the particle spends within $\calU$) rather than the local time. Nevertheless, one can derive a BVP for the associated propagator and incorporate partial absorption using a stopping occupation time distribution. Stochastic resetting can then be incorporated into the BVP along analogous lines to this paper. (The specific case of constant reactivity was developed in Ref. \cite{Schumm21}.)

\end{document}